\documentclass[10pt,final,conference,letterpaper,oneside]{IEEEtran}
\usepackage{cite}

\usepackage{graphicx}
\usepackage[caption=false,font=footnotesize]{subfig}

\usepackage{amsmath}
\usepackage{algorithmic}

\usepackage{array}

\ifCLASSOPTIONcompsoc
  \usepackage[caption=false,font=normalsize,labelfont=sf,textfont=sf]{subfig}
\else
  \usepackage[caption=false,font=footnotesize]{subfig}
\fi
\usepackage{url}

\usepackage{color}
\usepackage{ifthen}
\usepackage[normalem]{ulem}

\begin{document}
%
\title{E3Solver: decision tree unification by enumeration}



%
\author{\IEEEauthorblockN
	{M. Ammar Ben Khadra}
\IEEEauthorblockA{Dept. of Electrical and Computer Engineering\\
University of Kaiserslautern, Germany\\
khadra@eit.uni-kl.de}}

\IEEEspecialpapernotice{(Competition contribution)}

\maketitle

\begin{abstract}
We introduce E3Solver, a unification-based solver for programming-by-example (PBE) participating in the 2017 edition of the SyGuS Competition. 
Our tool proceeds in two phases. First, for each individual example, we enumerate a terminal expression consistent with it. Then, we unify these expressions using conditional expressions in a decision tree. 
To this end, a suitable condition is enumerated for each pair of \textit{conflicting} examples.
This incremental method terminates after fitting all examples in the decision tree. 
E3Solver solves all 750 instances in the bitvector sub-track in an average time of few seconds each.
We make our contributions publicly available (\url{https://github.com/sygus-tools})
\end{abstract}


%
\IEEEpeerreviewmaketitle

\section{Introduction}

Given syntactic constraints in the form of context-free grammar, and semantic constraints specified by logic formulas, a program sketch, or simply a set of I/O examples, the Syntax Guided Synthesis (\textsf{SyGuS}) problem is to find a function that satisfies the given syntactic and semantic constraints.

Enumerative search is a conceptually simple strategy for solving \textsf{SyGuS} problems were all possible expressions are evaluated in increasing order. Despite its relative simplicity, it has been shown in \cite{Udupa2013, Alur2013} that enumerative search, with some optimizations, can be effective especially for problems of smaller size. Note that if a solution is found using enumerative search, it's is provably the smallest solution possible.

The tool ESolver is an open source instantiation for the enumerative search strategy. In order to be more scalable, it implements the following key optimizations:

\textbf{Counter-Example Guided Inductive Synthesis (CEGIS)}. Instead of issuing a symbolic query to the SMT solver to validate each newly generated candidate expression, ESolver utilizes the counter-examples returned by the SMT solver in order to \textit{concretely} evaluate candidate expressions. 
Consequently, it queries the SMT solver only in case a candidate expression is consistent with all concrete counter-examples previously collected .

\textbf{Distinguishability pruning}. Validating all possible expression permutations is both unnecessary and expensive. For example,  consider the following language,

\begin{table}[h!]
	\vspace{-0.1cm}
	\centering
	\label{lbl:language}
	\begin{tabular}{r  l}
		$ term ~:=$      &  $false ~|~ true ~|~ x ~| ~ term  \wedge term $ \\
	\end{tabular}
	\vspace{-0.1cm}
\end{table}

Expressions $ false \wedge true$ and $x \wedge false $ are both indistinguishable w.r.t. expression $ false \wedge false$, i.e., they always 
evaluate to the same value $ false$. Therefore, ESolver tries to avoid generating indistinguishable expressions as much as possible. To this end, it maintains a data store of unique \textit{signatures} of previously evaluated expressions. In case the signature of a newly generated expression is already available in the store, this expression will be flagged as indistinguishable and won't be generated later as a subexpression. 

\section{Contributions}

We contribute two versions of E3Solver. The \textit{basic} version consists of maintenance and bug fixes to ESolver which made it capable of solving several programming-by-example (PBE) tasks. In the \textit{improved} version, we implemented our decision tree unification method for PBE. 
Both versions are available in the author's directory on StarExec platform. We are participating in the competition with the improved version only (\textbf{E3Solver}). Our main target is the bitvector sub-track.

\subsection{Basic version}

PBE track was introduced in the 2016 edition of SyGuS competition. Out of 750 PBE tasks in the bit-vector sub-track, ESolver was not capable of solving any. Implementation bugs contributed significantly to this situation.
For example, tasks in the \texttt{icfp} subfolder (450 in total)  were unparsable since they have Windows line endings (CRLF). 
Fixing this required a small modification to the \texttt{synthlib2parser}.

Additionally, relying on CEGIS triggers an exception in the program as the SMT solver repeatedly returns the same concrete example. 
Note that CEGIS is, in principle, not needed in PBE since the concrete values
of input variables are already available. 
That is, PBE tasks can be validated concretely without using an SMT solver. 

In this basic version, we automatically detect that the given constraints represent a PBE task and collect the concrete values of input variables accordingly. 
Solving proceeds then by enumeration exactly as in ESolver.

\subsection{Improved version}

Our decision tree unification method is implemented in this version. 
Basically, solving proceeds in two phases. First, we enumerate to find a terminal expression (\texttt{expr}) consistent with each example individually. 
Then, we incrementally build a decision tree by unifying pairs of examples. A pair of example is unified by inserting a decision node that branches between their terminal \texttt{expr}.
In this regard, our method shares similarities with decision tree learning \cite{Alur2017}. However, we build the decision tree using enumeration only. Now we discuss these phases in more detail.

\subsubsection{Building terminal expressions}

We collect the constraint that represents each individual example. 
This constraint takes the following form,

\begin{equation*}
[\bigwedge_{j = 0}^{n } (v_{j} = i_{j}) \wedge (f(v_{0}, .. v_{n}) = v_{t})] \Rightarrow (v_{t} = o)
\end{equation*}

\noindent where $f$ is the function to be synthesized. Inputs, outputs, and variables are represented by $i$, $o$, and $v$ respectively. 

For each example, we use enumerative search to find a valid expression (\texttt{expr}) consistent with its constraint. 
The latter \texttt{expr} will be checked against all examples that are not yet mapped to a valid expression. 
The intuition here is that multiple examples might exercise the same (or similar) control-flow path in $f$. Therefore, we do not need to start enumeration from scratch for each example.

\subsubsection{Building the decision tree}

This steps starts by picking two \textit{conflicting } examples in the previous step.
We say that a pair of examples are conflicting if they were mapped to different terminal \texttt{expr}. 
Then, these examples are unified by enumerating a condition expression. 
The latter expression must be (1) non-constant and (2) evaluates to 1 for the inputs 
of only one among the conflicting examples.

Later, we iterate over all remaining examples. For each example, we traverse the decision tree by evaluating the example's inputs against the condition of each traversed node. 
This depth-first traversal continues until we find a leaf decision node where we can insert the current example. Now, a new decision node is created such that it unifies the current example with the one mapped to the same branch in the leaf node.

The decision tree can grow in size to have a number of nodes in the order of $2^n$ where $n$ is the number of examples. 
We apply two optimizations in an attempt to reduce that. 
First, we rank the examples in a greedy manner such that examples with unique \texttt{expr} are put first in the queue to be unified. 
Second, we lazily expand the decision tree. To this end,  examples that share the same terminal \texttt{expr} and the same branch at a leaf node will not be immediately unified in a newly created decision node.

\section{Evaluation}

We experimented with both versions of E3Solver on the StarExec platform. Timeout was set to 1800 seconds and memory limit to 128 GB. The basic E3Solver solves 138 tasks in the bitvector sub-track out of 750 in total. 
The improved version solves all 750 tasks with an arithmetic mean of 11.8s and a median of 0.12s. 
Finally, note that E3Solver can act as a drop-in replacement for ESolver in the general track where it can solve all PBE tasks included in those benchmarks.

\section{Caveats}

In the current implementation, we assume that (1) $f$ is unary (has arity of one), and (2) there exist a grammar rule for a conditional expression. 
This rule must be named ``\texttt{if0}". 
These assumptions hold in the current benchmark set.


\bibliographystyle{IEEEtran}
\bibliography{sygus17ref}
%

\end{document}